\documentclass[prb,twocolumn,superscriptaddress,showpacs]{revtex4-1}

\usepackage{graphicx}
\usepackage{dcolumn}
\usepackage{bm}
\usepackage{amsmath}

\begin{document}

\preprint{APS/123-QED}

\title{Spin correlation and entanglement detection in Cooper pair splitters by current measurements using magnetic detectors}

\author{Piotr Busz}
\affiliation{Institute of Molecular Physics, Polish Academy of Science, 60-179 Poznan, Poland}
\author{Damian Tomaszewski}
\affiliation{Institute of Molecular Physics, Polish Academy of Science, 60-179 Poznan, Poland}
\author{Jan Martinek}
\affiliation{Institute of Molecular Physics, Polish Academy of Science, 60-179 Poznan, Poland}

\begin{abstract}
We analyze a model of double quantum dot Cooper pair splitter coupled to two ferromagnetic detectors
and demonstrate the possibility of determination of spin correlation by current measurements.
We use perturbation theory, taking account of the exchange interaction with the detectors,
which leads to complex spin dynamics in the dots.
This affects the measured spin and restricts the use of ferromagnetic detectors to the nonlinear current-voltage characteristic regime at the current plateau,
where the relevant spin projection is conserved,
in contrast to the linear current-voltage characteristic regime, in which the spin information is distorted.
Moreover, we show that for separable states the spin correlation
can only be determined in a limited parameter regime,
much more restricted than in the case of entangled states.
We propose an entanglement test based on the Bell inequality.
\end{abstract}


\maketitle

\section{Introduction }
Pairs of entangled particles provide the basis for modern applications in quantum cryptography, teleportation, and other topics in quantum information technology and quantum computation. Cooper pairs that naturally occur in the ground state of \textit{s}-wave superconductors provide a continuous solid-state source of spatially separated spin-entangled electrons, which can be used as flying qubits in integrated and scalable on-chip quantum information systems. A substantial breakthrough has been achieved recently in the theoretical modeling of Cooper pair splitting (CPS)~\cite{PhysRevB.63.165314,Lesovik2001} and in experimental realizations~\cite{Hofstetter2009,PhysRevLett.104.026801,Wei2010,PhysRevLett.107.136801,Das2012,PhysRevLett.109.157002,PhysRevLett.114.096602} by the introduction of a double quantum dot (DQD), soon followed by the attainment of a splitting efficiency close to 1.
An important step following a successful splitting of Cooper pairs is to verify experimentally whether the split electrons remain entangled. This turns out to represent a much greater challenge, as eight years after the first demonstration of Cooper pair splitting~\cite{Hofstetter2009} entanglement detection is still lacking in this system. Indeed, it is very difficult to find a suitable measurement scheme that would be both effective and relatively simple to realize experimentally.

\begin{figure}[t!]
\centering
\includegraphics{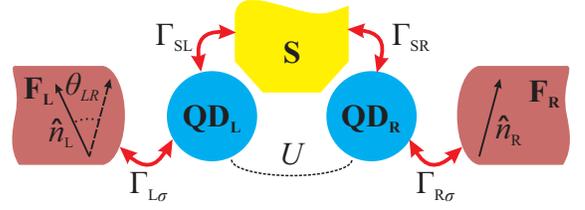}
\caption{(Color online) Schematic representation of the studied system,
with a superconducting electrode~\rm{S},
two quantum dots~$\rm{QD}_\eta$,
and two ferromagnetic electrodes~$\rm{F}_\eta$
with noncollinear magnetization directions~${\hat n_\eta}$,
where $\eta=\rm{L},\rm{R}$.}
\label{fig:uklad1}
\end{figure}

Most of the proposed verification methods~\cite{PhysRevLett.84.1035,Kawabata2001,PhysRevB.66.161320,PhysRevLett.91.157002,PhysRevLett.94.210601,PhysRevB.74.115315,Morten2008,PhysRevB.83.125304,PhysRevLett.118.036804} require the use of spin-sensitive detectors and higher-order cumulants, complex time-resolved measurements, or transfer of the spin state onto the polarization state of a pair of optical photons~\cite{PhysRevB.91.094516}, which are rather difficult experimental techniques. Some potentially simpler techniques based on dc current measurements~\cite{Soller2013,PhysRevLett.111.136806} were proposed on theoretical grounds in the last years. However, some of these proposals~\cite{Kawabata2001,Soller2013} neglect important physical aspects of the model, such as the Coulomb interaction, necessary to obtain a high splitting efficiency, or the exchange field-induced back action of the ferromagnetic detectors on the spin dynamics of the quantum dots, which can affect the results.
An interesting recent proposal, also based on dc current measurements and the spin-orbit interaction, involves the use of a bent carbon nanotube CPS under strong magnetic field~\cite{PhysRevLett.111.136806}. However, this technique has a disadvantage of using strong magnetic field, which can possibly modify the properties of the investigated ground state (as discussed in Ref.~[\onlinecite{PhysRevB.90.220501}]) and interfere with the measurements.
To avoid these difficulties we propose and analyze a perfectly natural setup for entanglement detection in CPS, in the form of noncollinear ferromagnetic spin detectors attached to both QDs~(Fig.~\ref{fig:uklad1}). This solution is experimentally feasible now~\cite{Pasupathy2004,f4fef170bebc11dcbee902004c4f4f50,Crisan2016} and has an additional advantage of involving simpler dc current measurements.

We develop a formalism~\cite{PhysRevLett.90.166602,PhysRevB.70.195345,PhysRevB.71.195324,PhysRevB.74.075328,PhysRevB.82.184507,Busz2015,PhysRevB.91.235424,arxiv} that represents a systematic approach taking into account the spin dynamics in the QDs and the exchange interaction between the ferromagnetic leads and the QDs~\cite{PhysRevLett.91.127203,Pasupathy2004,PhysRevB.72.121302,f4fef170bebc11dcbee902004c4f4f50,PhysRevLett.108.166605,Crisan2016}, issues not discussed in previous studies~\cite{PhysRevLett.84.1035,PhysRevB.66.161320,PhysRevLett.91.157002,PhysRevLett.94.210601,PhysRevB.74.115315,
Morten2008,PhysRevB.83.125304,Kawabata2001,Soller2013,PhysRevLett.111.136806,PhysRevLett.118.036804}. We prove that the complex spin dynamics in the QDs does not prevent the extraction of spin information, since the measured spin projection is conserved during spin precession in the nonlinear current-voltage regime at characteristic dc current plateaus (see Fig.~\ref{fig:Iapp(V)}). It is in contrast to the linear regime, where it is distorted, which has been ignored in previous studies. We demonstrate that the spin correlation function can be determined by dc current measurements at current plateaus in the nonlinear regime only. The spin correlation functions contain all the information necessary for the determination of the properties of the investigated ground state; therefore, using them we are able to test the Clauser-Horne-Shimony-Holt (CHSH) Bell inequalities to discriminate between entangled and unentangled product states. We analyze the limitations of the entanglement detection scheme based on ferromagnetic detectors attached to the CPS and its sensitivity to various asymmetries.
\section{The model }
The Hamiltonian~$H$ of the considered three-terminal system is defined as:
\begin{equation}
H = {H_{{\rm{DQD}}}} + \sum\limits_{\eta  = {\rm{L}},{\rm{R}}} {\left( {{H_\eta } + {H_{{\rm{T}}\eta }}} \right)}  + {H_{\rm{S}}} + {H_{{\rm{TS}}}}.
\end{equation}
The first term is related to the single-level double quantum dot (DQD):
\begin{align}
{H_{{\rm{DQD}}}} =& \sum\limits_{\eta ,\sigma  =  \uparrow , \downarrow } { {{\varepsilon _{\eta \sigma }}{n_{\eta \sigma }}}  + \sum\limits_\eta  {{U_\eta }\,{n_{\eta  \uparrow }}{n_{\eta  \downarrow }}} }\nonumber \\
&  + U\sum\limits_{\sigma ,\sigma '} {{n_{{\rm{L}}\sigma }}} {n_{{\rm{R}}\sigma '}}\;,
\label{eqn:ddot}
\end{align}
where ${n_{\eta\sigma }}$  is the number operator of particles
in the QD~$\eta=\rm{L}/\rm{R}$ (left/right) with spin $\sigma$, energy ${{\varepsilon _{\eta \sigma }}}$,  and $U$ is the Coulomb interaction between the two QDs.
The intradot Coulomb repulsion $U_\eta$ is infinite,
which means that the dot can only be occupied by a single electron.

The ferromagnetic metal electrodes, acting as spin detectors, are treated as reservoirs of noninteracting fermions with momentum $k$ and spin $\alpha$:
\begin{align}
{H_{\eta}} =&\sum\limits_{k,\alpha  =  \uparrow , \downarrow } {{\varepsilon _{\eta k}}} a_{k\eta\alpha }^\dag {a_{k\eta\alpha }}\;.
\label{eqn:heta}
\end{align}
The effective spin asymmetry, ${\rho _{\eta \uparrow }} \ne {\rho _{\eta \downarrow }}$, in the density of states ${\rho _{\eta \alpha }}$ at Fermi level in the electrodes can be described by spin polarization
\mbox{${p_\eta } = \left( {{\rho _{\eta  \uparrow }} - {\rho _{\eta  \downarrow }}} \right)/\left( {{\rho _{\eta  \uparrow }} + {\rho _{\eta  \downarrow }}} \right)$}.
In general, the magnetization directions ${\hat n_{\rm{L}} }$ and ${\hat n_{\rm{R}} }$
of the left and right  leads, respectively,
are noncollinear, ${\hat n_{\rm{L}}\neq\hat n_{\rm{R}} }$~\cite{PhysRevB.39.6995,Moodera1996,Crisan2016}.
To describe the spin conserving tunneling, taking account of rotation of the spin
quantization axes,
we have to include SU(2) rotation matrices $\hat U_{\alpha \sigma }^\eta$, with elements \mbox{$U_{\alpha \sigma }^\eta  = \left\langle {{{\eta\alpha }}}
 \mathrel{\left | {\vphantom {{{\eta\alpha }} {{\eta\sigma  }}}}
 \right. \kern-\nulldelimiterspace}
 {{{\eta\sigma }}} \right\rangle$}, into the tunneling Hamiltonian~\cite{PhysRevB.74.075328}:
 \begin{equation}
{ H_{\rm{T}\eta }} = \sum\limits_{k,\eta,\sigma ,\alpha  } {\left( {{V_{\eta}}a_{k\eta\alpha }^\dag U_{\alpha \sigma }^\eta {d_{\eta\sigma }} + {\rm{H}}.{\rm{c}}.} \right)}\;,
\label{eqn:tunnF}
\end{equation}
where $V_{\eta}$ denotes the tunneling amplitude
between QD~$\eta$ and ferromagnetic lead~$\eta$;
$d_{\eta\sigma }$ and $a_{k\eta\alpha }$ are the annihilation operators in QDs and leads, respectively.

The superconducting lead can be described by the mean-field BCS Hamiltonian:
\begin{align}
{H_{\rm{S}}} =&\sum\limits_{k,\sigma  =  \uparrow , \downarrow } {{\varepsilon _{\rm{S}k}}} a_{k\rm{S}\sigma }^\dag {a_{k\rm{S}\sigma }}\nonumber \\
&- \Delta \sum\limits_k \left({a_{- k\rm{S} \downarrow }^\dag {a_{k\rm{S}\uparrow }^\dag} + \,} \rm{H}.\rm{c}.\right)\;,
\label{eqn:sc}
\end{align}
where $\Delta $ is the pair potential and a reference electrochemical potential~${\mu _{\rm{S}}} = 0$. The BCS Hamiltonian~(\ref{eqn:sc}) yields an \textit{s}-wave superconductor, where each Cooper pair is in a spin singlet ground state~$|\rm{S}\rangle$,
that can be generalized to an arbitrary ground state of Cooper pair
given by: $|\varphi \rangle  \equiv {a_1}\left| {{ \uparrow _{\rm{L}}}{ \uparrow _{\rm{R}}}} \right\rangle  + {a_2}\left| {{ \uparrow _{\rm{L}}}{ \downarrow _{\rm{R}}}} \right\rangle  + {a_3}\left| {{ \downarrow _{\rm{L}}}{ \uparrow _{\rm{R}}}} \right\rangle  + {a_4}\left| {{ \downarrow _{\rm{L}}}{ \downarrow _{\rm{R}}}} \right\rangle $,
where $\sum\limits {{|a_{\rm{j}}|^2} = 1} $. The tunneling between the superconducting electrode and QD~${\eta }$ is given by:
\begin{equation}
{ H_{\rm{TS}}} = \sum_{k\eta\sigma} {\left( {{V_{\rm{S}\eta }}a_{k\rm{S}\sigma }^\dag {d_{\eta\sigma }} + {\rm{H}}.{\rm{c}}.} \right)}\;.
\end{equation}
The tunnel coupling strengths to the two ferromagnetic electrodes are expressed as
${\Gamma _{\eta  \uparrow / \downarrow }} = \left( {1 \pm {p_\eta }} \right){\Gamma _\eta }/2$,
where ${\Gamma _\eta } = 2\pi \left( {{\rho _{\eta  \uparrow }} + {\rho _{\eta  \downarrow }}} \right)|{V_\eta }{|^2}$, and to the superconducting one as $\Gamma_{\rm{S}\eta} = 2 \pi \rho_{\rm{S}}   | V_{\rm{S}\eta}|^2$, where $\rho_{\rm{S}}$ denotes
density of states in superconducting leads.

In our study we consider transport processes involving Andreev reflection.
By tracing out the degrees of freedom of the superconducting electrode
we obtain the effective Hamiltonian of the DQD
that takes account of the coupling to the superconducting lead:
\begin{align}
{H_{{\rm{eff}}}} =& {H_{{\rm{DQD}}}} - \frac{{{\Gamma _{\rm{S}}}}}{{\sqrt 2 }}\left( {a_1}d_{{\rm{L}} \uparrow }^\dag d_{{\rm{R}} \uparrow }^\dag  + {a_2}d_{{\rm{L}} \uparrow }^\dag d_{{\rm{R}} \downarrow }^{\dag}\right.\nonumber\\
&\left.+ {a_3}d_{{\rm{L}} \downarrow }^\dag d_{{\rm{R}} \uparrow }^\dag  + {a_4}d_{{\rm{L}} \downarrow }^\dag d_{{\rm{R}} \downarrow }^\dag  + {\rm{H}}.{\rm{c}}. \right)\;,
\label{eqn:eff}
\end{align}
where $\Gamma_{\rm{S}}= \sqrt{\Gamma_{\rm{SL}}\Gamma_{\rm{SR}}}$
and the second term describes the nonlocal proximity effect. The diagonalization of Hamiltonian~(\ref{eqn:eff}) yields the coupling between two states, the empty state~$|0\rangle$ and the two-particle state~$|\varphi \rangle$,
which results in the new
Andreev bound
eigenstates:
\begin{equation}
| \pm \rangle  = {w_ \mp }|0\rangle  \mp {w_ \pm }|\varphi \rangle\;,
\end{equation}
with amplitudes ${w_ \mp } = \sqrt {1/2 \mp \delta /4{\varepsilon _{\rm{A}}}} $,
where $\delta=\varepsilon_{\rm{L}}+\varepsilon_{\rm{R}} +U$ denotes the detuning parameter.
The energies of states $| \pm \rangle $ in the diagonal basis are
$ E_{\pm}=\delta/2\pm \epsilon_{\rm{A}}$, where \mbox{$\epsilon_{\rm{A}} = \sqrt{\delta^2/4 + \Gamma _{\rm{S}}^2/2}$}.
If we consider singlet pairing in the superconductor $\left| \varphi  \right\rangle  =\left| \rm{S} \right\rangle $, then ${a_2} =  - {a_3} = 1/\sqrt 2 $ and ${a_1} = {a_4} = 0$.
\section{Master equations}
We consider spin-dependent electron transport to the lowest order in~$\Gamma_\eta$,
a regime known as the sequential tunneling limit, $\Delta  \gg {k_{\rm{B}}}T>\Gamma_\eta$,
easily reachable in current experiments, which allows us neglect quasiparticle excitations in the superconductor.
The net tunneling rate to and from ferromagnetic lead $\eta$ depends on the direction of lead magnetization,
which we describe by spinors~\cite{PhysRevB.71.195324} $m_{\eta \uparrow }^\dag  = \left( {U_{ \uparrow \uparrow }^{\eta \; * },U_{ \uparrow \downarrow }^{\eta \; * }} \right)$ and $m_{\eta \downarrow }^\dag  = \left( {U_{ \downarrow \uparrow }^{\eta \; * },U_{ \downarrow \downarrow }^{\eta \; * }} \right)$.

Let us discuss in detail the formalism for~$\left| \varphi  \right\rangle = \left| \rm{S} \right\rangle$. The restriction~$\mu <\left({\delta  + U}\right)/{2}$ for symmetric  bias voltages
$\left({\mu _{\rm{L}}} = {\mu _{\rm{R}}} = \mu \right)$
allows us to neglect triplet states~\cite{comm1}.
Thus, we only consider six states: two
states~$\left|  \pm  \right\rangle $
with occupancy probabilities $p_\pm $,
and four
single-electron states~$\left| {\eta\sigma } \right\rangle$
described by density matrices
\mbox{${\rho _{1\eta }} = \left({{p_{1\eta }}}/{2}\right) {\rm{I}} + {S_{\rm{X}\eta }}{ \sigma _{\rm{X}}} + {S_{\rm{Y}\eta }}{ \sigma _{\rm{Y}}} + {S_{\rm{Z}\eta }}{ \sigma _{\rm{Z}}}$},
where ${{p_{1\eta }}}$ denotes the probability of the single electron occupancy of the QD~$\eta$,
${ \vec{S}_\eta } = \left({S_{\rm{X}\eta }},{S_{\rm{Y}\eta }},{S_{\rm{Z}\eta }}\right)$ is the average spin vector in QD~$\eta$,
and $\vec{\sigma} = \left( {\sigma} _{\rm{X}},\sigma _{\rm{Y}}, \sigma _{\rm{Z}}\right)$ is the Pauli matrix vector. A quantum dot state is characterized by a set of ten parameters, $\left\{{p_ + },{p_ - },
{p_{\rm{1L}}},{p_{\rm{1R}}},{S_{\rm{XL}}},{S_{\rm{YL}}},{S_{\rm{ZL}}},{S_{\rm{XR}}},{S_{\rm{YR}}},{S_{\rm{ZR}}}\right\}$.
Due to the normalization condition $1~ = {p_ - } + {p_ + } +\sum_{\eta}p_{1\eta }$ only nine of them are independent.

The time evolution of the scalars~$p_\pm $ and density matrices~${\rho _{1\eta }}$
is described by the following effective master rate equations:
\begin{align}
{\hbar }\frac{{d{\rho _{1\eta }}}}{{dt}} =& \frac{i}{\hbar }{\Big[{\rho _{1\eta }},{H_{1\eta }}\Big]_ - } + \sum\limits_{\sigma ,\rm{s} = \{  + , - \} } \bigg( f_{\eta \sigma } ^{ + \rm{s} - }{\gamma _{\rm{\bar s}\eta \sigma}}{p_{\rm{s}}} \gamma _{\rm{\bar s}\eta \sigma }^\dag \nonumber\\
& + f_{\bar \eta\sigma }^{ - \rm{s} + }{\bar \gamma _{\rm{s}\bar \eta \sigma}}{p_{\rm{s}}}\bar \gamma _{\rm{s}\bar \eta \sigma}^\dag  - \frac{1}{2}{}f_{\eta \sigma } ^{ - \rm{s} - }{\left[{\gamma _{\rm{\bar s}\eta \sigma}}\gamma _{\rm{\bar s}\eta \sigma}^\dag ,{\rho _{1\eta }}\right]_ + }\nonumber\\
&- \frac{1}{2}f_{\bar \eta\sigma }^{ + \rm{s} + }{\left[{\bar \gamma _{\rm{s}\bar \eta \sigma}}\bar \gamma _{\rm{s}\bar \eta \sigma}^\dag ,{\rho _{1\eta }}\right]_ + }\bigg)\;,
\nonumber\\
{\hbar }\frac{{d{p_ \mp }}}{{dt}} =&\sum\limits_{\sigma,\eta}  \Big( -  f_{\eta \sigma }^{ +  \mp  - }\gamma _{\pm\eta \sigma }^\dag {p_ \mp }{\gamma _{\pm\eta \sigma}}\nonumber\\
& - f_{\bar \eta\sigma }^{ -  \mp  + }\bar \gamma _{\mp\bar \eta \sigma}^\dag {p_ \mp }{\bar \gamma _{\mp\bar \eta \sigma}}
+ f_{\eta \sigma } ^{ -  \mp  - }\gamma _{\pm\eta \sigma}^\dag {\rho _{1\eta }}{\gamma _{\pm\eta \sigma}} \nonumber\\
& + f_{\bar \eta\sigma }^{ +  \mp  + }\bar \gamma _{\mp\bar \eta \sigma}^\dag {\rho _{1\eta }}{\bar \gamma _{\mp\bar \eta \sigma}}\Big)\;,
\label{eqn:pmp2}
\end{align}
where the square brackets ${[\;]_ \mp }$ denote the commutator/anticommutator,
and the tunneling amplitude spinor ${\gamma _{{\rm{s}}\eta \sigma}}={\left| {{\omega _{\rm{s}}}} \right|}\Gamma _{\eta \sigma }^{1/2}{m_{\eta \sigma }}$, ${\bar \gamma _{{\rm{s}}\eta \sigma }} = \left({1}/{{\sqrt 2 }}\right){\left| {{\omega _{\rm{s}}}} \right|}\Gamma _{\eta \sigma }^{1/2}{m_{\eta \bar\sigma }}$. The additional factor ${1}/{{\sqrt 2 }}$ in ${\bar \gamma _{s\eta \sigma }}$ is a consequence of the participation of singlet state $|S\rangle $ in the given process. The symbols $\bar \sigma $ and $\bar\eta$ denote the spin opposite to $\sigma$ and the ferromagnetic lead opposite to $\eta$, respectively. We use spinor ${\gamma _{{\rm{s}}\eta \sigma }}$ for the description of processes changing the occupancy of the DQD
between empty and single,
and ${\bar \gamma _{{\rm{s}}\eta \sigma }}$ for processes switching between single and double. The tunneling amplitude spinor ${\bar \gamma _{s\eta \sigma }}$ can indicate the entanglement of a singlet state.
The tunneling of one electron, with spin $\sigma$, of an $|S\rangle $ pair from QD~$\eta$ to ferromagnetic electrode~$\eta$ causes the collapse of the two-particle wave function; thus the next Cooper pair electron
in QD~$\bar\eta$ has the opposite spin, described by spinor ${m_{\eta \bar\sigma }}$.
A similar effect occurs in electron tunneling in the opposite direction, i.e., from ferromagnetic electrode~$\eta$ to QD~$\eta$. In the adopted formalism we use the following notation for the Fermi distribution functions ${f_{\eta}^ {+} }\left(\zeta \right)$:
$f_{\eta \sigma }^{ \pm  {\rm{s}} + } = f_{\eta \sigma }^  \pm  \left({E_{\rm{s}}} - {\varepsilon _{\bar \eta \bar \sigma }}\right)$,
$f_{\eta \sigma }^{ \pm {\rm{s}} - } = f_{\eta \sigma }^  \pm  \left({\varepsilon _{\eta \sigma }} - {E_{\rm{s}}}\right)$,
where ${f^ - } = 1 - {f^ + }$ and the third subscript indicates the change in the DQD occupation,
$+$ between double and single and $-$ between single and empty.

The Hamiltonian
\begin{align}
{H_{1\eta }} =& \frac{\hbar }{{4\pi }}{\rm{P}}\int d \xi \sum\limits_{\sigma ,{\rm{s}} \in \{ {\rm{ + }}, - \} }  {\frac{{{ \left(1 - 2f_{\eta \sigma }^ + \left(\xi \right)\right)\gamma _{{\rm{\bar s}}\eta \sigma }}\gamma _{{\rm{\bar s}}\eta \sigma }^\dag }}{{{\varepsilon _{\eta \sigma }} - {E_{\rm{s}}} - \xi }}}\;
\label{eqn:h11eta}
\end{align}
describes virtual particle exchange processes resulting in an effective exchange field \mbox{${\vec{B}_\eta }$}
and spin precession for a single electron states $\left| {\eta\sigma } \right\rangle$ around the direction of ${\vec{B}_\eta }$. Here $\textrm{P}$ denotes the Cauchy principal value. Using the relation ${S_{{\rm{i}}\eta }} = \left( {1/2} \right){\rm{Tr}}\left[ {{\rho _{{\rm{1}}\eta }}{\kern 1pt} {\sigma _{\rm{i}}}} \right]$ $\left({\rm{i}} \in \left\{ \rm{X},\rm{Y},\rm{Z}\right\}\right) $ it can be shown that the expression $d{\rho _{1\eta }}/dt = \left( {i/\hbar } \right){\left[ {\rho _{1\eta }},{H_{1\eta }}\right]_ - }$ is equivalent to the Bloch equation:
\begin{align}
d{\vec S_\eta }/dt = {\rm{ }}{\vec S_\eta } \times {\vec B_\eta }\;,
\label{eqn:bloch}
\end{align}
which describes the spin precession around the effective field \mbox{${\vec{B}_\eta } = {B_\eta }{\hat n_\eta }$}, where
\begin{align}
{B_\eta } =& \frac{{{\Gamma _{\eta  \downarrow }} - {\Gamma _{\eta  \uparrow }}}}{{\hbar \pi }}\rm{P}\int d \xi \sum\limits_{{\rm{s}} \in \{ {\rm{ + }}, - \} } \frac{{\left| {{\omega _{\bar{\rm{s}}}}} \right|}^{2}{\it{f}_{\eta }^ - \left(\xi \right)}}{{{\varepsilon _{\eta }} - {E_{\rm{s}}} - \xi }}\;.
\label{eqn:amplituda}
\end{align}
The effective exchange field results not only in a torque of the accumulated spin,
but also in a spin splitting of the dot level, similar to Zeeman splitting~\cite{PhysRevLett.91.127203,Pasupathy2004,PhysRevB.72.121302,f4fef170bebc11dcbee902004c4f4f50,PhysRevLett.108.166605}. In the weak-coupling regime it
cannot be resolved, since the splitting is proportional to the coupling strength and must be dropped in first-order transport calculation~\cite{PhysRevLett.90.166602,PhysRevB.70.195345}.
In the case of two-electron states $\left| \varphi  \right\rangle $, spin splitting and precession can be neglected
due to weak coupling to ferromagnetic electrodes, ${\Gamma _{\rm{S}\eta }} \gg {\Gamma _\eta }$, in the considered system.

The current in electrode~$\eta$ is given by the following equation:
\begin{align}
{I_\eta } =&\frac{e}{\hbar } \sum\limits_{{\rm{s}},\sigma } {\Big(f_{\eta \sigma } ^{ + {\rm{s}} - }\gamma _{{\rm{\bar s}}\eta \sigma }^\dag {p_{\rm{s}}}{\gamma _{{\rm{\bar s}}\eta \sigma }}}  - f_{\eta \sigma } ^{ - {\rm{s}} - }\gamma _{{\rm{\bar s}}\eta \sigma }^\dag {\rho _{1\eta }}{\gamma _{{\rm{\bar s}}\eta \sigma }}\nonumber\\
&- f_{\eta \sigma } ^{ - {\rm{s}} + }\bar \gamma _{{\rm{s}}\eta \sigma }^\dag {p_{\rm{s}}}{\bar \gamma _{{\rm{s}}\eta \sigma }} + f_{\eta \sigma } ^{ + {\rm{s}} + }\bar \gamma _{{\rm{s}}\eta \sigma }^\dag {\rho _{1\bar \eta }}{\bar \gamma _{{\rm{s}}\eta \sigma }\Big)}\;,
\label{eqn:cur}
\end{align}
where occupancy probabilities can be obtained from the stationary solution of Eq.~(\ref{eqn:pmp2}).
The studied model provides 100\% efficiency of Cooper pair splitting;
thus, $I_{\rm{L}}=I_{\rm{R}}$ and the total current ${I} ={I_{\rm{L}}}+{I_{\rm{R}}}$.
In the case of nonmagnetic electrodes \mbox{($p_\eta=0$)} our model is equivalent to that presented in Ref.~[\onlinecite{PhysRevB.82.184507}].
In particular magnetization configurations we can denote the total current as~$I_{\uparrow_{\rm{L}}\uparrow_{\rm{R}}}$,
where $\uparrow_\eta$ indicates~$\hat n_\eta$, while $\downarrow_\eta$ indicates~$-\hat n_\eta$.
For example, $I_{\uparrow_{\rm{L}}\downarrow_{\rm{R}}}$ describes a reversal
of the magnetization direction of electrode~R, ${\hat n_{\rm{R}}} \Rightarrow  - {\hat n_{\rm{R}}}$. This implies ${m_{{\rm{R}} \uparrow }} \Rightarrow {m_{R \downarrow }}$ and ${m_{R \downarrow }} \Rightarrow-{m_{R \uparrow }}$. The other configurations, ${I_{{ \downarrow _{\rm{L}}}{ \uparrow _R}}}$ and ${I_{{ \downarrow _{\rm{L}}}{ \downarrow _{\rm{R}}}}}$,
are defined similarly.
Our aim is to determine the spin correlation of the ground state $\left| \varphi  \right\rangle $ of the superconductor
by the measurement of spin-dependent currents in different configurations of the electrode magnetizations.
We seek evidence that Cooper pairs that occupy the QDs
are still in a quantum entangled state for $\left| \varphi  \right\rangle  = \left| \rm{S} \right\rangle $.
\section{Collinear configurations}
\begin{figure}[t!]
\centering
\includegraphics{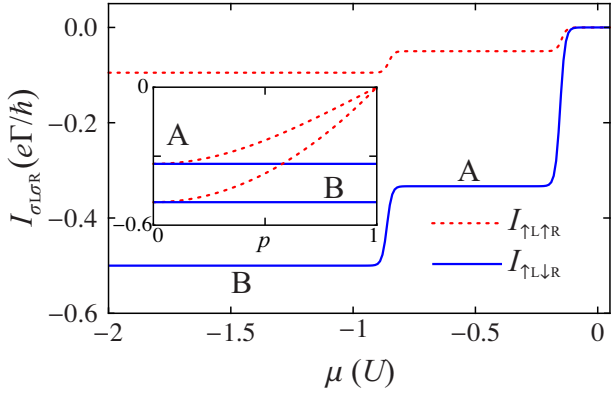}
\caption{(Color online) Currents~${I_{{\uparrow _{\rm{L}}}{\uparrow _{\rm{R}}}}}$~(dotted line) and ${I_{{{\uparrow }_{\rm{L}}}{\downarrow{\rm{R}}}}}$~(solid line)
versus voltage~$\mu$
for collinear parallel and antiparallel magnetizations, respectively,
of the ferromagnetic leads
in a symmetric system:
$\Gamma_{\rm{L}} = \Gamma_{\rm{R}}=\Gamma$, $\epsilon_{\rm{L}} = \epsilon_{\rm{R}}$, $p_L=p_R=p=0.9$, $\delta=0$,
$k_{\rm{B}}T=0.01{\kern 1pt} U$, and $\Gamma_S=0.5{\kern 1pt} U$.
In the inset, ${I_{{\uparrow _{\rm{L}}}{\uparrow _{\rm{R}}}}}$ and ${I_{{{\uparrow }_{\rm{L}}}{\downarrow{\rm{R}}}}}$,
versus the spin polarization~$p$ for two plateaus~(A,~B).}
\label{fig:Iapp(V)}
\end{figure}
Let us first consider the currents in the case of
collinear magnetizations of the ferromagnetic electrodes, \mbox{${\hat n_{\rm{L}}} = \pm {\hat n_{\rm{R}}}$}.
Currents~${I_{{\sigma _{\rm{L}}}{{ \sigma }_{\rm{R}}}}}$ and ${I_{{\sigma _{\rm{L}}}{{\bar \sigma }_{\rm{R}}}}}$
for parallel (P) and antiparallel (AP) magnetizations, respectively, are plotted versus voltage $\mu$ in Fig.~\ref{fig:Iapp(V)}.
Two characteristic plateaus are observed in the plot:
(A) with $\left| {\eta \sigma } \right\rangle$ and $\left|  -  \right\rangle$ as the only states participating in transport,
and (B) with state $\left|  +  \right\rangle $ available as well (see Ref.~[\onlinecite{PhysRevB.82.184507}] for details).
Close to $\mu=0$ only states $\left| {\eta \sigma } \right\rangle $ are occupied
and the system is in the Coulomb blockade regime.
The P~configuration current ${I_{{\sigma _{\rm{L}}}{\sigma _{\rm{R}}}}}$
is much smaller than the AP~configuration current ${I_{{\sigma _{\rm{L}}}{{\bar \sigma }_{\rm{R}}}}}$
and decreases with increasing spin polarization~$p_L=p_R=p$, as shown in Fig.~\ref{fig:Iapp(V)}, inset.
The AP configuration current~${I_{{{ \sigma }_{\rm{L}}}{\bar\sigma _{\rm{R}}}}}$ is independent of~$p$
and equal to the current~$I_0$ in a system with nonmagnetic electrodes ($p=0$),
${I_{{{\sigma }_{\rm{L}}}{\bar\sigma _{\rm{R}}}}}=I_{0}$.
This is related to the fact that the Cooper pairs are in singlet states~$|\rm{S}\rangle$
and the AP alignment of electrode magnetizations better suits the antiferromagnetic order of the singlet state.
We can try to use this sensitivity of the current to magnetization configuration to determine
the spin correlation of the Cooper pairs directly from electric current measurements.
\section{Spin dynamics }
\begin{figure}[h!]
\centering
\includegraphics{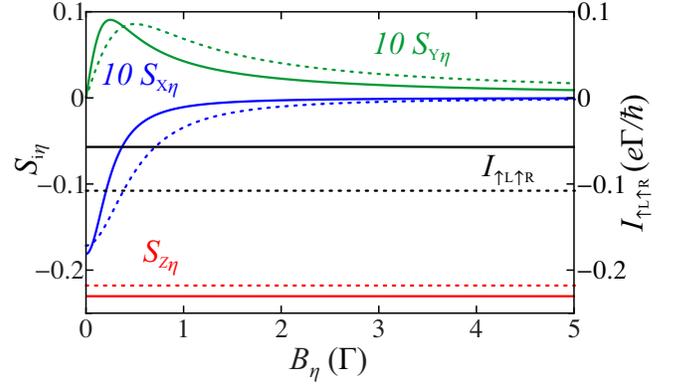}
\caption{(Color online) Current $I_{ \uparrow_{\rm{L}}\uparrow_{\rm{R}} }$ flowing to the ferromagnetic electrodes and Cartesian components of electron spin in a quantum dot $\eta$ versus the exchange field $B_\eta$ associated with the ferromagnetic electrode $\eta$ (independent of $B_{\bar\eta})$. The results are obtained for a symmetrical system with an \textit{s}-wave superconducting electrode at the two plateaus (A~-~solid lines, B~-~dotted lines), for $\hat n_{\eta}$ along the $Z$ direction, the angle between directions of electrode magnetizations $\theta_{{\rm{LR}}}=\pi/4$, $p=0.9$, and $k_BT=0.01\,U.$}
\label{fig:sxyzi}
\end{figure}
\begin{figure}[t!]
\centering
\includegraphics{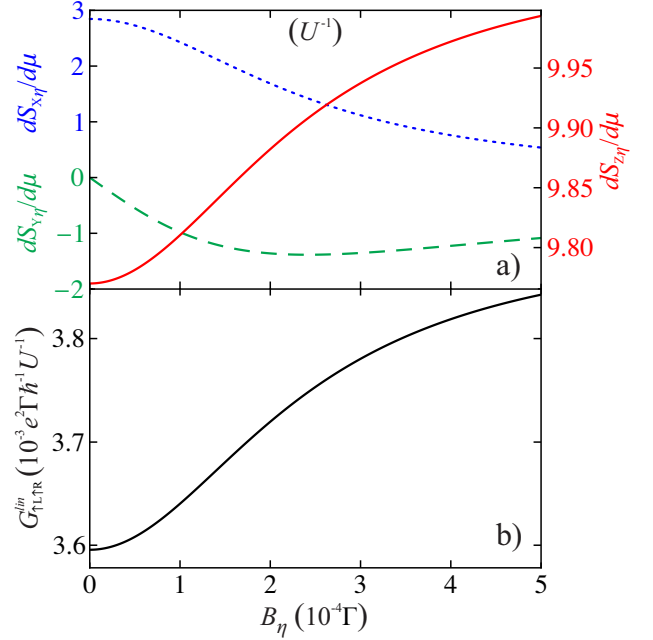}
\caption{(Color online) (a) Cartesian components ($S_{\rm{Z}\eta}$~-~solid, $S_{\rm{Y}\eta}$~-~dashed, $S_{\rm{X}\eta}$~-~dotted line) of electron spin in a quantum dot $\eta$, and (b) linear differential conductance ${G_{ \uparrow_{\rm{L}}\uparrow_{\rm{R}} }^{lin}}$ versus the exchange field $B_\eta$ associated with ferromagnetic electrode $\eta$ (independent of $B_{\bar\eta})$. Results obtained for a symmetrical system with an \textit{s}-wave superconducting electrode in the linear response regime. Here $\hat n_{\eta}$ is along the $Z$ direction, and the angle between the directions of electrode magnetizations is $\theta_{{\rm{LR}}}=\pi/4$, $p=0.9$, and $k_BT=0.02\,U.$}
\label{fig:sxyzg}
\end{figure}
In this paper we propose a method for entanglement detection using noncollinear ferromagnetic electrodes as effective spin detectors.
However, such noncollinear ferromagnetic detectors can affect the state of a quantum dot ~\cite{PhysRevLett.90.166602,PhysRevB.70.195345,PhysRevB.71.195324} and possibly distort the acquired spin information.
The presence of the ferromagnetic electrodes in the considered system results in a complex spin dynamics in the QDs~\cite{PhysRevLett.90.166602,PhysRevB.70.195345,PhysRevB.71.195324}.
Virtual particle exchange processes between ferromagnetic electrode~$\eta$ ($\eta=\{L,R\}$) and QD~$\eta$~\cite{PhysRevLett.91.127203,Pasupathy2004,f4fef170bebc11dcbee902004c4f4f50,PhysRevLett.108.166605,Crisan2016} lead to an effective exchange field~${\vec{B}_\eta }$ [Eq.~(\ref{eqn:amplituda})], which in the sequential tunneling limit causes precession of electron spin in the QDs.
This spin precession can be described by the Bloch equation Eq.~(\ref{eqn:bloch}).

Fortunately, the complex spin dynamics induced by the field~$\vec{B}_\eta $ associated with the ferromagnetic electrodes
does not interfere with the reading of the spin detectors.
This rather surprising effect is one of the important results of our study,
since it allows us to determine the spin correlation
by measuring the current~${I_{ \sigma_{\rm{L}}  \sigma_{\rm{R}} }}$.
Figure~\ref{fig:sxyzi} shows that the current ${I_{{\sigma _{\rm{L}}}{\sigma _{\rm{R}}}}}$ at both plateaus~(A,~B)
is independent of the field~$\vec{B}_\eta $.
This is because the spin ${\vec S_\eta }$ in QD~$\eta$ precesses around the direction of $\vec{B}_\eta$,
parallel to the magnetization direction $\hat n_\eta $ of electrode~$\eta$,
and its projection $S_{\rm{Z}\eta}$ on this direction is conserved (Fig.~\ref{fig:sxyzi}).
Thus, the quantity relevant to the measurement is not affected,
in contrast to the spin components $S_{\rm{Y}\eta}$, perpendicular to the plane spanned by the two magnetizations, and $S_{\rm{X}\eta}$,
which do change with the field amplitude~$B_\eta$.

In the linear response regime, $\mu  \ll {k_B}T$, the differential conductance~${G_{ \uparrow_{\rm{L}}\uparrow_{\rm{R}} }^{lin}} = {\left. {e(\partial I_{ \uparrow_{\rm{L}}\uparrow_{\rm{R}} }/\partial \mu )} \right|_{\mu  = 0}}$,
plotted versus~$B_\eta$ in Fig.~\ref{fig:sxyzg}(b),
is affected by the spin precession, which excludes the use of ferromagnetic leads
as spin detectors in this limit.
As shown in Fig.~\ref{fig:sxyzg}(a), apart from $S_{\rm{X}\eta}$ and $S_{\rm{Y}\eta}$,
also the spin component $S_{\rm{Z}\eta}$ in the direction $\hat n_\eta $
varies with the exchange field~$B_{\eta}$.
This is caused by the possibility of tunneling in the reverse direction, from ferromagnetic electrodes to quantum dots.

The plots for the nonlinear (on plateaus), Fig.~\ref{fig:sxyzi}, and linear response regimes, Fig.~\ref{fig:sxyzg}, have different scales of exchange field~$\vec{B}_\eta $.
In the linear response regime the electron spin in a QD is more sensitive to the field~$\vec{B}_\eta $,
because of the lower current and the related longer dwell time~\cite{Braun2005},
the average time spent by an electron on the quantum dot.
Therefore, saturation is observed for much lower values of~$B_{\eta}$ with respect to the plateaus.

We prove that at the observed current plateaus the field~$\vec{B}_\eta $ does not interfere with the reading of spin in the electrodes.
Thus, the measured current~${I_{ \sigma_{\rm{L}}  \sigma_{\rm{R}} }}$
(where $\sigma_\eta$ indicates the magnetization direction~$\hat n_\eta$ of ferromagnetic electrode~$\eta$)
can be used for determining spin correlations and testing the Bell inequalities.
\section{Spin correlation}
Since the current ${I_{{\sigma _{\rm{L}}}{\sigma _{\rm{R}}}}}$ depends on the magnetization direction,
we can try to extract spin information from it.
The two-spin correlation~$C_{{\rm{LR}}}^{\rho}$
can be calculated from the equation
$C_{{\rm{LR}}}^{\rho}=\rm{Tr}[\left( {{\sigma _{\rm{L}}} \otimes {\sigma _{\rm{R}}}} \right)\rho ]$,
where ${\sigma _{{\rm{L}}/{\rm{R}}}} = \vec \sigma \cdot{\hat n_{{\rm{L}}/{\rm{R}}}}$ and $\rho\equiv\left| \varphi  \right\rangle \left\langle \varphi  \right|$ denotes the two-particle density matrix.

To extract spin information from the direct current we propose the following function:
\begin{equation}
C_{{\rm{LR}}}^I = \frac{{{I_{{ \uparrow _{\rm{L}}}{ \uparrow _{\rm{R}}}}} + {I_{{ \downarrow _{\rm{L}}}{ \downarrow _{\rm{R}}}}} - {I_{{ \uparrow _{\rm{L}}}{ \downarrow _{\rm{R}}}}} - {I_{{ \downarrow _{\rm{L}}}{ \uparrow _{\rm{R}}}}}}}{{{I_{{ \uparrow _{\rm{L}}}{ \uparrow _{\rm{R}}}}} + {I_{{ \downarrow _{\rm{L}}}{ \downarrow _{\rm{R}}}}} + {I_{{ \uparrow _{\rm{L}}}{ \downarrow _{\rm{R}}}}} + {I_{{ \downarrow _{\rm{L}}}{ \uparrow _{\rm{R}}}}}}}\;,
\label{eqn:col}
\end{equation}
and test its correspondence to the spin-spin correlation function~$C_{{\rm{LR}}}^{\rho}$.
Equation.~(\ref{eqn:col}) is analogous to the correlation function defined for the number of coincidences~\cite{PhysRevB.89.125404}.

Let us first consider the symmetric case
(i.e. $p_{\rm{L}} = p_{\rm{R}}=p$, ${{\Gamma _{\rm{L}}} = {\Gamma _{\rm{R}}}}=\Gamma$, and $\delta=0$).
In this regime we are able to reproduce the spin correlations for any state $\left| \varphi  \right\rangle $
up to a spin polarization-dependent amplitude~$\Im (p)$:
\begin{equation}
C_{{\rm{LR}}}^I = \Im (p)\,C_{{\rm{LR}}}^\rho\;.
\label{eqn:rówcol}
\end{equation}
This, however, can only be done at the two plateaus, in the symmetric case $\Im (p) = 2{p^2}/(3 - {p^2})$ and $\Im (p) = {p^2}/(2 - {p^2})$
for plateaus A and B, respectively~\cite{simple}.
This is main result of our study, since by studying~$C_{{\rm{LR}}}^I $ from Eq.~(\ref{eqn:rówcol}) we can obtain information on the spin correlations in our system,
and detect entanglement of split
Cooper pairs.

\begin{figure}[t!]
\centering
\includegraphics{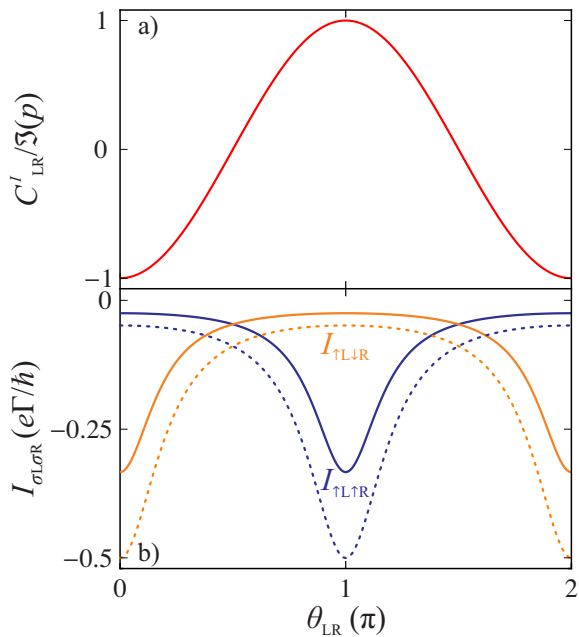}
\caption{(Color online) (a) Spin correlation function~$C_{{\rm{LR}}}^I$ for the singlet state $\left| \varphi  \right\rangle=\left| S  \right\rangle $, (b) currents ${I_{ \uparrow_{\rm{L}}  \uparrow_{\rm{R}} }}$ and ${I_{ \uparrow_{\rm{L}}  \downarrow_{\rm{R}} }}$ for two configurations of ferromagnetic electrode magnetizations, plotted versus
the angle~$\theta_{\rm{LR}}$ between the magnetization directions~$\hat n_\eta$
in a symmetric system for $p=0.95$ at two plateaus~(A~-~solid,~B~-~dotted lines).}
\label{fig:coliI}
\end{figure}

 Figure~\ref{fig:coliI}(a) shows the spin correlation~$C_{{\rm{LR}}}^I$
as a function of the angle~$\theta_{{\rm{LR}}}$ between the electrode magnetization directions
for a singlet state \mbox{$\left| \varphi  \right\rangle  = \left| \rm{S} \right\rangle $}.
Interestingly, although the current~${I_{{\sigma _{\rm{L}}}{\sigma _{\rm{R}}}}}(\theta_{{\rm{LR}}})$,
plotted in Fig.~\ref{fig:coliI}(b), does not follow simple
${N_{{\sigma _{\rm{L}}}{\sigma _R}}}$ coincidence predictions,
the spin correlator $C_{\rm{LR}}^I(\theta_{\rm{LR}}) =-\Im (p) \cos (\theta_{{\rm{LR}}})$ behaves as predicted by quantum theory $C_{\rm{LR}}^\rho(\theta_{\rm{LR}}) =-\cos (\theta_{{\rm{LR}}})$.
The $\theta_{{\rm{LR}}}$ and $p$ dependence of the currents ${I_{{\sigma _{\rm{L}}}{\sigma _{\rm{R}}}}}$ in Fig.~\ref{fig:coliI}(b) is described by the following expressions:
\begin{align}
&{I_{ \uparrow_{\rm{L}}  \uparrow_{\rm{R}} }} =\frac{{{p^2} - 1}}{{2{p^2}\cos (\theta_{{\rm{LR}}} ) - {p^2} + 3}}\;,\nonumber\\
&{I_{ \uparrow_{\rm{L}}  \downarrow_R }} = \frac{{1 - {p^2}}}{{2{p^2}\cos (\theta_{{\rm{LR}}} ) + {p^2} - 3}}\;,
\label{eqn:cur2}
\end{align}
and
\begin{align}
&{I_{ \uparrow_{\rm{L}}  \uparrow_{\rm{R}} }} = \frac{{{p^2} - 1}}{{{p^2}\cos (\theta_{{\rm{LR}}} ) - {p^2} + 2}}\;,\nonumber\\
&{I_{ \uparrow_{\rm{L}}  \downarrow_{\rm{R}} }} = \frac{{1 - {p^2}}}{{{p^2}\cos (\theta_{{\rm{LR}}} ) + {p^2} - 2}}\;,
\label{eqn:cur3}
\end{align}
for plateaus A and B, respectively.

\begin{figure}[h!]
\centering
\includegraphics{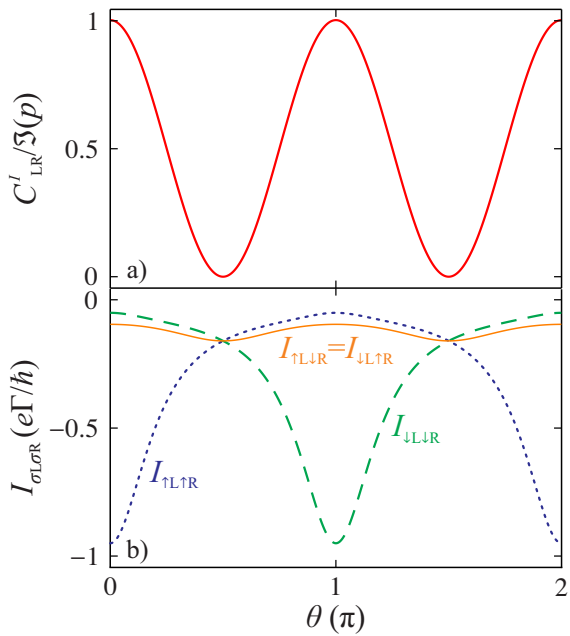}
\caption{(Color online) (a) Spin correlation function~$C_{{\rm{LR}}}^I$ for the product state $\left| \varphi  \right\rangle $ and (b) currents~${I_{{\uparrow _{\rm{L}}}{\uparrow _{\rm{R}}}}}$-~dotted, ${I_{{\uparrow _{\rm{L}}}{\downarrow_{\rm{R}}}}}$-~solid, and ${I_{{\downarrow_{\rm{L}}}{\downarrow_{\rm{R}}}}}$-~dashed line, for three configurations of the magnetic leads  in the symmetric system at the plateau B, plotted versus angle~$\theta$ for $\alpha_{{\rm{L}}}=-\alpha_{{\rm{R}}}=\theta$, $p=0.9$, and $k_BT=0.01\,U.$}
\label{fig:icsep}
\end{figure}
We can use our model for the determination of the spin correlation of electrons
in a Cooper pair naturally occurring in the superconductor.
In general, the ground state~$\left| \varphi  \right\rangle $ of the superconductor
can be an entangled state or a separable state~\cite{RevModPhys.75.657}.
The essential difference between these two kinds of quantum states
is that, in contrast to an entangled state,
particles in a separable state are independent of each other. In the case of separable pure states (product states)
each dot in our system has a well-defined spin in the state~$|\varphi \rangle$.
In a two-spin product state the spin correlation function~\mbox{$C_{{\rm{LR}}}^\rho=\cos(\alpha_{{\rm{L}}})\cos(\alpha_{{\rm{R}}})$}
depends only on the angles $\alpha_{{\rm{\eta}}}$ ($\eta=\{\textrm{L},\textrm{R}\}$)
between the magnetization direction~$\hat n_{\eta}$ and the spin direction
at QD~$\eta$ in the state~$|\varphi \rangle$.
In a symmetric system
(i.e., for $p_{\rm{L}} = p_{\rm{R}}=p$, ${{\Gamma _{\rm{L}}} = {\Gamma _{\rm{R}}}}$ and $\delta=0$)
we can also determine the spin correlation for separable states in the DQD
by measuring the current,~Eq.~(\ref{eqn:rówcol}).

Let us consider the case when $\alpha_{{\rm{L}}}=-\alpha_{{\rm{R}}}=\theta$.
The spin correlation function $C_{{\rm{LR}}}^I$ obtained by measurement of the current Eq.~(\ref{eqn:col})
is plotted versus $\theta$ in Fig.~\ref{fig:icsep}(a).
Shown in Fig.~\ref{fig:icsep}(b), the currents ${I_{{\sigma _{\rm{L}}}{\sigma _{\rm{R}}}}}(\theta)$ and ${I_{{\sigma _{\rm{L}}}{\bar\sigma _{\rm{R}}}}}(\theta)$,
for parallel and antiparallel magnetizations, respectively,
do not follow simple
${N_{{\sigma _{\rm{L}}}{\sigma _R}}}$ coincidence predictions,
but the spin correlator~$C_{{\rm{LR}}}^I=\Im(p) {\kern 1pt}\cos ^2(\theta )$ behaves as predicted by quantum theory $C_{{\rm{LR}}}^\rho=\cos ^2(\theta )$.
\section{Entanglement detection by testing Bell inequalities}

The Bell inequalities~\cite{RevModPhys.38.447,Physics} concern measurements of separated particles that interacted before the separation.
Assuming a local realism~\cite{PhysRev.47.777,PhysRev.85.166}, certain constraints must hold on the relationships
between the correlations between successive measurements of the particles in various possible measurement settings.
We use the CHSH version of the Bell inequality~\cite{PhysRevLett.23.880} with the correlator $Q^\rho$ defined by means of spin correlation functions as:
\begin{equation}
{Q^\rho } = \left| {C_{{\rm{LR}}}^\rho  + C_{{\rm{L'R}}}^\rho  + C_{{\rm{LR'}}}^\rho  - C_{{\rm{L'R'}}}^\rho } \right| \le 2\;.
\label{eqn:chsh}
\end{equation}
When the inequality~(\ref{eqn:chsh}) is not fulfilled the particles are in an entangled state.

Using the fact that we can determine the spin correlation by the current measurements, Eq.~(\ref{eqn:rówcol}), we can test the Bell inequality using the current measurements as well.
The CHSH correlator in our system has the following form:
\begin{equation}
Q^{I} = \Im (p){Q^\rho }\;,
\label{eqn:rówQ}
\end{equation}
where we have substituted $C_{{\rm{LR}}}^\rho$ in Eq.~(\ref{eqn:chsh}) with $C_{{\rm{LR}}}^I$ defined by Eq.~(\ref{eqn:col}).

For a singlet state~$|\rm{S}\rangle $ the inequality~(\ref{eqn:chsh}) is maximally violated for example,
when \mbox{$\theta_{{\rm{L}}',{\rm{R}}}=\theta_{{\rm{R}},{\rm{L}}}=\theta_{{\rm{L}},{\rm{R}}'}=\pi /{4}$} and $\theta_{{\rm{L}}',{\rm{R}}'}=3\pi /{4}$;
then, ${Q^\rho }=2\sqrt 2$.

In the next step we determine the system parameter limits within which the Bell inequality can be violated
by a singlet state $\left| \varphi  \right\rangle  = \left| \rm{S} \right\rangle $.
We obtain the minimum spin polarization~$p$ of the leads necessary for the violation of the CHSH inequality for plateaus A and B, $p > \sqrt {\frac{3}{7}(2\sqrt 2  - 1)}\approx0.885$ and $p > \sqrt {2\sqrt 2  - 2}\approx0.91$, respectively.
\section{Results for asymmetric system}
\begin{figure}[t!]
\centering
\includegraphics{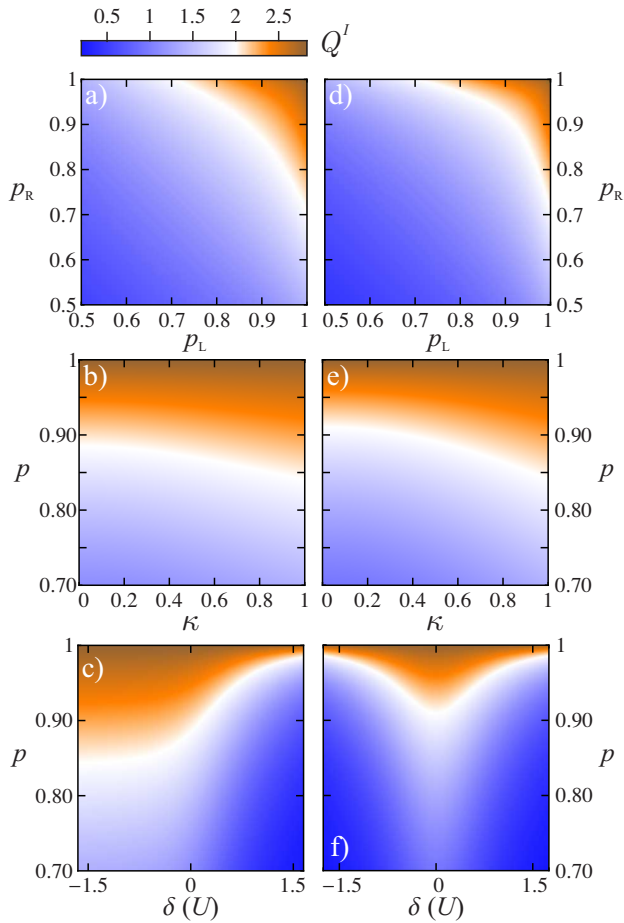}
\caption{(Color online) Density plot of the CHSH correlator~$Q^{I}$
for a system with a superconducting electrode~$|S\rangle $ as a function of:
(a, d) spin polarizations~$(p_{\rm{L}},p_{\rm{R}})$ of the leads for symmetric coupling,
(b, e) spin polarization~$p_{\rm{L}}=p_{\rm{R}}=p$ of the leads and coupling asymmetry~$\kappa$,
and (c, f) spin polarization~$p$ of the leads and detuning parameter~$\delta$ for
symmetric coupling. All the results are obtained for $Q^{\rho}=2\sqrt{2}$ at plateau~A (a-c) and at plateau~B (d-f); $k_BT=0.01\,U$.
Orange areas represent ranges in which entangled state can be detected.}
\label{fig:zaku05}
\end{figure}
Now let us consider a case with a number of asymmetries:
$p_{\rm{L}} \neq p_{\rm{R}}$, $\kappa  = (\Gamma _{\rm{L}} - \Gamma _{\rm{R}})/(\Gamma _{\rm{L}} + \Gamma _{\rm{R}})\neq0$, $\delta\neq0$,
and some difference between the dot energy levels, \mbox{$\Delta \varepsilon=\varepsilon_{\rm{L}}-\varepsilon_{\rm{R}} $}.
We find that the quantity~$C_{\rm{LR}}^I$ defined in Eq.~(\ref{eqn:col})
still describes the spin correlation, $C_{\rm{LR}}^I = \Im \,C_{\rm{LR}}^\rho$, for all maximally entangled Bell states: $\left| {{\Psi ^ \pm }} \right\rangle  \equiv\frac{1}{{\sqrt 2 }}\left( {\left| {{ \uparrow _{\rm{L}}}{ \downarrow _{\rm{R}}}} \right\rangle  \pm \left| {{ \downarrow _{\rm{L}}}{ \uparrow _{\rm{R}}}} \right\rangle } \right)$ and $ \left| {{\Phi ^ \pm }} \right\rangle  \equiv \frac{1}{{\sqrt 2 }}\left( {\left| {{ \uparrow _{\rm{L}}}{ \uparrow _{\rm{R}}}} \right\rangle  \pm \left| {{ \downarrow _{\rm{L}}}{ \downarrow _{\rm{R}}}} \right\rangle } \right)$.

The correlation function~$C_{{\rm{LR}}}^I$ does not depend on~$\Delta \varepsilon $,
and the other asymmetries mentioned above only affect the amplitude~$\Im $ of the correlator.
For \mbox{$p_{\rm{L}} \neq p_{\rm{R}}$}, \mbox{$\kappa=0$} and \mbox{$\delta=0$},
knowing that \mbox{$\Im (p_{\rm{L}},p_{\rm{R}})>{\sqrt 2 }/{2}$},
we can specify the conditions to be fulfilled by the spin polarizations of the leads $(p_{\rm{L}},p_{\rm{R}})$ for quantum entanglement
to be detected. Figure~\ref{fig:zaku05}(a) shows the range of $p_{\rm{L}}$ and $p_{\rm{R}}$ (orange area)
where entanglement can be detected at the characteristic current plateau~A.
Figure~\ref{fig:zaku05}(b) presents
the CHSH correlator~$Q^{I}$ versus spin polarization~$p$ and coupling asymmetry~$\kappa$ at plateau~A
for a system with ${{\Gamma _{\rm{L}}}\neq{\Gamma _{\rm{R}}}}$ as the only asymmetry.
The applicability range of the proposed method is found to grow with~$\kappa$. For the extreme value $\kappa=1$ the requirement for spin polarization $p$ is minimal,
and corresponds to the condition~$p > {1}/{{\sqrt[4]{2}}}$~$\approx0.84$ established in Refs.~[\!\!\citenum{PhysRevB.89.125404,Roek2015}].
Analogous results at plateau~B are in Figs.~\ref{fig:zaku05}(d) and ~\ref{fig:zaku05}(e). The parameters range in which entanglement detection is possible at plateau~B is smaller with respect to plateau~A. The influence of the detuning parameter~$\delta $ on the detection of state~$|S\rangle $
has a different character for the plateau~A [Fig.~\ref{fig:zaku05}(c)] and plateau~B [Fig.~\ref{fig:zaku05}(f)].

Unfortunately, for separable states~$\left| \varphi  \right\rangle$
in an asymmetric system, $C_{{\rm{LR}}}^I \neq \Im {\kern 1pt}  C_{{\rm{LR}}}^\rho$,
which implies distorted spin information.
To exclude that inequality~(\ref{eqn:chsh}) might be unfulfilled by separable states
we have analyzed the corresponding CHSH correlator~$Q^{I}$,
and found that separable states can only violate the CHSH inequality
in a very restricted range of parameters in a strongly asymmetric DQD system
when the spin polarization~$p_\eta$ in one of the leads is close to 1, $(1-p_{\eta})\lesssim{3\times10^{ - 3}}$;
this, however, is difficult to achieve experimentally.
Thus, beyond this restricted regime the correlator~$Q^{I}$ can be used
for detecting maximally entangled states.

\section{Conclusions}

We have studied theoretically the role of ferromagnetic electrodes
connected to two QDs of a CPS to act as spin detectors
converting spin information directly into a charge current.
We have derived effective master equations describing transport in the system
with the exchange interaction and the related spin dynamics taken into account.
Despite the complexity of the spin dynamics,
the conservation of the relevant spin projection allows for
the determination of spin correlations from current measurements
and the detection of entanglement by testing the Bell inequality.
The spin correlation of maximally entangled states is insensitive to various asymmetries in the system parameters.
In the case of separable states, symmetry conservation is required for the determination of the spin correlation.

We would like to thank J.~Barna\'{s}, A.~Bednorz, W.~Belzig, M.~Braun, B.~Braunecker, F.~Dominguez, T.~Kontos, J.~K\"onig,
C.~Sch\"{o}nenberger, B.~Sothmann and A.~L.~Yeyati for helpful discussions.
This study has received support from the EU~FP7 Project SE2ND (No.~271554) and the National Science Centre of Poland, Grant No.~2015/17/B/ST3/02799.

\end{document}